\documentclass[a4paper,conference]{IEEEtran}
\usepackage{hyperref}
\usepackage{comment}
\usepackage{graphicx}
\usepackage[caption=false]{subfig}
\usepackage{booktabs}
\usepackage{graphicx}
\usepackage{color}
\usepackage{bm}
\usepackage{tikz} 
\usetikzlibrary{arrows,decorations.pathmorphing,backgrounds,fit,positioning,shapes.symbols,chains}
\tikzset{
    box1/.style={%
        draw=black, thick,
        rectangle,
        minimum height=3cm,
        minimum width=4cm
    },
}

\definecolor{myblue}{RGB}{91,155,213}
\renewcommand{\vec}{\bm}

\begin{document}
\title{Detecting and adapting to crisis pattern with \\
context based Deep Reinforcement Learning}

\author{
\IEEEauthorblockN{Eric Benhamou$^{1}$, \hspace{0.5cm} David Saltiel$^{2,3}$, \hspace{0.5cm}  Jean-Jacques Ohana$^{4}$, \hspace{0.5cm} and Jamal Atif$^{1}$}
\IEEEauthorblockA{$^1$MILES, Machine Learning Group, LAMSADE, Dauphine, France\hspace{0.2cm}   $^2$Ai Square Connect, Research group, France}
\IEEEauthorblockA{$^3$LISIC, ULCO, France\hspace{0.2cm}   $^4$Multi Assets Solutions, Homa Capital, France}
\IEEEauthorblockA{$^1$Email: \{eric.benhamou, jamal.atif\}@lamsade.dauphine.fr, $^2$Email: david.saltiel@aisquareconnect.com}
\IEEEauthorblockA{$^4$Email: jjohana@homacapital.fr}
}


%
\maketitle              
\begin{abstract}
Deep reinforcement learning (DRL) has reached super human levels in complex tasks like game solving (Go \cite{silver2017mastering}, StarCraft II \cite{Vinyals_2019}),  and autonomous driving \cite{Wang2018DeepRL}. 
However, it remains an open question whether DRL can reach human level in applications to financial problems and in
particular in detecting pattern crisis and consequently dis-investing.
In this paper, we present an innovative DRL framework consisting in two sub-networks fed respectively
with portfolio strategies past performances and standard deviations as well as additional contextual features. 
The second sub network plays an important role as it captures dependencies with common financial indicators features like risk aversion, economic surprise index and correlations between assets that allows taking into account context based information. We compare different network architectures either using layers of convolutions to reduce network's complexity or LSTM block to capture time dependency and whether previous allocations is important in the modeling. We also use adversarial training to make the final model more robust.
Results on test set show this approach substantially over-performs traditional portfolio optimization methods like Markovitz and is able to detect and anticipate crisis like the current Covid one.
\end{abstract}

\section{Introduction}\label{sec:intro}
Being able to adapt portfolio allocation to crisis environment like the current Covid crisis is a major concern for the financial industry. Indeed, the current Covid crisis took the industry by surprise twice. First, when stock markets plunged at an unprecedented speed in March 2020 with the SP 500 falling by 13 \%, asset managers were slow to react and to cut risk exposure. And secondly, when stock markets bounced back up at an equally rapid pace, with a rise of 13 \% for  the SP 500 in May 2020, asset managers were again overhauled. In contrast, the previous 2008 crisis was very slow both in terms of its falls and recovery. Hence, adapting portfolio allocation to crisis environment is a very important matter and has attracted growing attention from the financial scientific community. 

The standard approach for portfolio allocation, that serves as a base line for our research, relies on determining portfolio weights according to a risk return criterion. The so called Markovitz portfolio \cite{markowitz1952portfolio} finds the optimal allocation by determining the portfolio with minimum variance given a target return or equivalently the portfolio with maximum return given a targeted level of variance (the dual optimization).  However, this approach suffers from a major flaw because of unreliable risk estimations of the individual portfolio strategy excess returns and covariances. This leads not only to unstable allocations, but also to slow reactions to changing environments \cite{black1992global}. If we want to find a more dynamic allocation method, deep reinforcement learning is an appealing method. It reformulates the portfolio optimization problem as a continuous control program with delayed rewards. Rules are simple. Each trading day, the dynamic virtual asset manager agent has the right to modify the portfolio allocation. When it modifies the portfolio weights, it incurred transaction costs. The agent can only allocate between 0 and 100 \% for all the portfolio assets. It can not short any asset, hence weights are always positive and never above 100 \%. It can not neither borrow to fund leverage positions, hence the sum of all allocations is strictly equal to 100 \%. To make decisions, the dynamic agent has access not only to past performances but also some financial contextual information that helps it making an informed decision. The agent receives in terms of feedback a financial reward that orientates its decisions. Compared to traditional financial methods, this approach has the major advantage to adapt to changing market conditions and to be somehow more model free than traditional financial methods as we connect portfolio allocations directly to financial data and not to specific risk factors that may factor in some cognitive bias. This stream of research is also highly motivated by the recent major progress of deep reinforcement learning methods that have reached super human levels in complex tasks like game solving (historically Atari games \cite{mnih-atari-2013}, Go \cite{silver2017mastering}, StarCraft II \cite{Vinyals_2019}),  and autonomous driving \cite{Wang2018DeepRL}. Nonetheless, it still remains an open question whether DRL can reach human level in applications to financial problems and in particular in detecting pattern crisis and consequently dis-investing.

\subsection{Related Work}
Initially, many of the machine-learning and in particular deep network applications to financial markets tried to predict price movements or 
trends \cite{Freitas_2009}, \cite{Niaki2013}, \cite{Heaton_2017}. The logic was to take historical prices of assets as inputs and use deep neural networks to predict asset prices for the next period.
Armed with the forecast, a trading agent can act and decide the best allocation. The problem to solve is a standard supervised learning task and more precisely a regression problem.  It is straightforward to implement. Yet, the efficiency of the method relies heavily on the accuracy of the prediction, which makes the method quite fragile and questionable as future market prices are well known to be difficult to predict. Furthermore, this approach tends to reduce substantially portfolio diversification and can not cope easily with transactions costs. In contrast, DRL can easily tackle these issues as it does not aim at predicting prices but rather at finding the optimal action or for our matter the optimal allocation. 

The idea of applying DRL to portfolio allocation has recently taking off in the machine learning community with some recent works on crypto currencies \cite{Jiang_2016}, \cite{Zhengyao_2017}, \cite{Liang_2018}, \cite{Yu_2019} and \cite{Wang_2019}.
Compared to traditional approaches on financial time series, that aim at taking decision based on forecasting estimates, \cite{Zhengyao_2017} and \cite{Liang_2018} showed that deep reinforcement learning with Convolutional Neural Network (CNN) architecture tends to perform better for crypto currencies and Chinese stock markets than deep learning architecture that relies on time series forecast like LSTM. However, when there is a very rapid crisis, like what happened during the Covid crisis, using just past performances may lead the DRL agent to react too slowly. 
To make an analogy, it is as if the agent was self-driving on the highway and very brutally, an obstacle arises. Using past performances only is like looking in the mirrors behind to infer what will happen next. Adding a context is like lifting up our eyes and looking further forward. 
Context based reinforcement learning has recently emerged as strong tool to increase reinforcement learning dynamic agent performance \cite{Gupta_2018,Lee_2020}.
More specifically, context based reinforcement learning (RL) with high capacity function approximators, such as deep neural networks (DNNs), has in the last two years attracted growing attention and been the subject of many publications in notorious machine learning conferences as it solved efficiently a variety of sequential decision-making problems, including board games (e.g., Go and Chess (\cite{Schrittwieser_2019}), video games (e.g., Atari games (\cite{Kaiser_2020}), and complex robotic control tasks (\cite{Zhang_2018}, \cite{Nagabandi_2019}, \cite{Hafner_2020}).  Theoretically, it has also been advocated that the usage of a context enables achieving superior data-efficiency to model-free RL methods in general \cite{Deisenroth_Rasmussen_2011}, \cite{Levine_Abbeel_2014}.

So in this work, we extend previous works of DRL by precisely using a context based approach. This is done by integrating common financial states in our deep network, having at least two sub networks and potentially three if we also incorporates in states the previous allocations. Experiments show that this approach is able to pick the best portfolio allocation out of sample using financial features used by asset managers: risk aversion index, correlation between equities and bonds, Citi economic surprise index and to accommodate for crisis by reducing risk exposure. We provide performances out of sample and test various configurations to emphasize that using CNN works much better than more predictive architecture like LSTM confirming previous works.

\subsection{Contributions}
Our contributions are twofold:
\begin{itemize}
\item First we explain why a context based deep reinforcement learning approach is closer to human thinking and leads to better results, with a novel deep network architecture consisting of two sub networks: one network (network 1) that takes as inputs past performances (and standard deviation) of the portfolio strategies and another one (network 2) that takes as inputs financial contextual information related to the performances of the portfolio strategies that are thought to have some predictive power regarding portfolio strategies future performances.
\item Second, we summarize lots of empirical findings. Reward function is critical. Sharpe ratio reward leads to different results compared to a straight final net performance reward function. CNN performs better than LSTM and captures implicit features. Using adversarial training by adding noise to the data improves the model. Last but not least, dependency to previous allocations does not improve the model.
\end{itemize}

\section{Mathematical formulation}
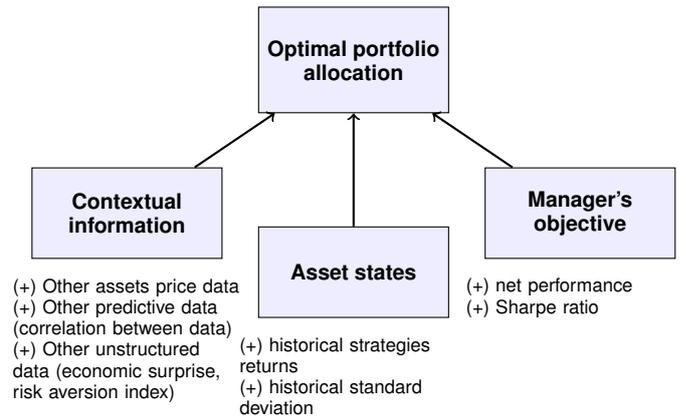
\begin{figure}[h]

\centering

\begin{tikzpicture}
[node distance = 1cm, auto,font=\footnotesize,
every node/.style={node distance=3cm},
comment/.style={rectangle, inner sep= 0pt, text width=3cm, node distance=0.25cm, font=\scriptsize\sffamily},
box/.style={rectangle, draw, 
fill=blue!70, fill opacity=0.1, 
text=black,text opacity=1,
inner sep=0pt, text width=2.5cm, text badly centered, minimum height=1.2cm, font=\bfseries\footnotesize\sffamily},
box2/.style={rectangle, draw,
fill=blue!70, fill opacity=0.1, 
text=black,text opacity=1,
inner sep=0pt, text width=2.5cm, text badly centered, minimum height=1.4cm, font=\bfseries\footnotesize\sffamily}] 

\node [box2] (allocation) {Optimal portfolio\\ allocation};
\node [below=1.2 cm of allocation](empty){};
\node [box, below=1.5 cm of allocation] (strategies) {Asset states};
\node [box, left=1.6cm of empty] (context) {Contextual \\ information};
\node [box, right=1.6cm of empty] (objective) {Manager's objective};

\node [comment, below=0.25 of strategies]
{
(+) historical strategies returns\\
(+) historical standard deviation
};

\node [comment, below=0.25cm of context] 
{
(+) Other assets price data\\
(+) Other predictive data (correlation between data)\\
(+) Other unstructured data (economic surprise, risk aversion index)
};

\node [comment, below=0.25cm of objective] 
{
(+) net performance\\
(+) Sharpe ratio\\
};


\path[->,thick] 
(strategies) edge (allocation)
(context) edge (allocation)
(objective) edge (allocation);

\end{tikzpicture} 
\caption{Portfolio allocation problem}
\label{fig:pap}
\end{figure}

As summarized by figure \ref{fig:pap}, an asset manager robot has several strategies that it wants to allocate optimally, with a performance objective on the overall portfolio. Not only does it have access to historical daily performance (the middle rectangle in figure  \ref{fig:pap}) but it can also leverage additional information (the rectangle on the left in figure  \ref{fig:pap}) that provides some contextual information about market conditions. These are other price data points but also unstructured data like some macro economic data. To gauge the performance of its decision, it has an objective that can be either the net performance of the portfolio or some risk return criterion (the third rectangle of figure  \ref{fig:pap} on the right)

The question of the asset allocation can be reformulated as a standard reinforcement learning problem, thanks to Markov Decision Process (MDP).  
The learning agent interacts with an environment $E$ to decide rational or optimal actions and receives in return some rewards.
These rewards are not necessarily only positive and are given only at the end of the financial episode. These rewards act as a feedback for finding the best action. 
Using the established formalism of Markov decision process, we assume that there exists a discrete time stochastic control process represented by a 4-tuple defined by  $(\mathcal{S}$, $\mathcal{A}$, $P_a, R_a)$ where $\mathcal{S}$ 
is the set of states, $\mathcal{A}$ the set of actions, 
$P_a(s, s') = \Pr(s_{t+1}=s' \mid s_t = s, a_t=a)$ the transition probability that action $a$ in state $s$ 
at time $t$ will lead to state $s'$ at the next period $t+1$ and finally, $R_a(s,a)$ the immediate reward received after state $s$ and action $a$. \\

The requirement of a Markovian state that guarantees that there exists a solution (hence satisfying the Bellman optimality principle \cite{SuttonBarto_2018}) is a strong assumption that is hard to verify in practice. 
It is somehow levied in practice by stacking enough observations to enforce that the Markov property is satisfied. Hence, it is useful, following \cite{Mnih_2016} or \cite{Jaderberg_2016}, to introduce the concept of observations and pile them to coin states. In this setting, the agent perceives at time $t$ an observation $o_t$ along with a reward $r_t$. \\

In our setting, time is divided into trading periods of equal length $\tau$. In the rest of the paper, $\tau$ represents one trading day but the setting can be applied to shorter time periods, like 30 minutes, to deal with intraday trading decisions. At the beginning of each trading period, a trading robot decides to potentially reallocate the funds among $m$ assets. 
The trading robot has access to an environment that provides at each time $t$, a state $s_t$ that is composed of the past observations that are rich enough to assume Markovianity.
Intuitively, it is important for the agent to observe not only the last returns but also some previous returns (like the returns over 2, 3 and 4  business days, but also a week and potentially a month) to make a decision.
Mathematically, we denote by $\delta_1$ the lag operator applied to each observation. To make this concrete the lag operator $\delta_1$ 's outputs are the last portfolio strategy returns at time $t$ but also at time $t-1,\ t-2,\ t-3$ and so on. There is here some trade-off. We obviously need enough observations to mimick a Markovian setting to ensure problem is well posed. 
But we also need to reduce observations to avoid facing the curse of dimensionality.
We will discuss this point in our experience, but practically, we take returns at time $t-60$ representing returns 3 months ago, $t-20$ one month ago\footnote{as there are approximately 60 trading days in a quarter, and 20 days in a month}, $t-4,\  t-3,\ t-2,\ t-1$ and $t$, the latter four providing returns over the last trading week.

By abuse of language, we can represent the lag $\delta_1$ operator by a vector of lagging periods $\delta_1 =  \left[ 0, 1, 2, 3 , 4, 20, 60 \right]$ (as there is a one to one mapping between the operator and the lagging periods) and retrieve the corresponding returns for asset $i$ as follows: $\left[r^i_t,r^i_{t-1},r^i_{t-2}, \ldots, r^i_{t-30},  r^i_{t-60} \right]$. Inputs that we call asset states as they directly relate to the portfolio's assets are not only past returns lagged over the $\delta_1$ periods but also standard deviation. The intuition behind the consumption of returns standard deviation or equivalently their volatilty is that volatility is a good predictor of crisis. Indeed it is a stylized fact in the financial literature that volatility is a good predictor of risk. \cite{Ross_1976} \cite{Harmon_2010} and that an increase of volatility comes swiftly after a market crash \cite{Black1976StudiesOS} \cite{Wu_2001}.
The period to compute the volatility is a hyper parameter, again another hyper-parameter to fine-tune and is arbitrarily taken to 20 periods to represent a month of data. If we summarize, asset states $A_t$ are given by two matrices $A_t = \left[ A^1_t, A^2_t \right]$ with the first matrix $A^1_t$ (in red)  being the the matrix of returns:\\

\begin{center}
\begin{tikzpicture}
[
box/.style={rectangle, draw, fill=red, fill opacity=0.1, color = red!80,
rectangle, minimum height=4cm, minimum width=6cm,
inner sep=0pt, text width=3cm, minimum height=2cm, font=\bfseries\footnotesize\sffamily},
comment2/.style={rectangle, inner sep= 0pt, text width=3cm, node distance=0.25cm, font=\sffamily}
]
\node[box, fill=red, fill opacity=0.1, color = red!80] (c2) at (0,0) {};
\node [comment2, left= -3 cm of c2, color = black ] {
$\begin{array}{l l l }
A^1_t = &
\left( 
\begin{array}{l l l }
r^1(t) 	& \ldots 	& r^1(t-60) \\
\ldots  	& \ldots 	& \ldots  \\
r^m(t) 	& \ldots 	& r^m(t-60)
\end{array} \right)
\end{array}
$
};
\end{tikzpicture}
\end{center}

while the second matrix (in blue) $ A^2$ containing standard deviations:

\begin{center}
\begin{tikzpicture}
[
box/.style={rectangle, draw, fill=red, fill opacity=0.1, color = red!80,
rectangle, minimum height=4cm, minimum width=6cm,
inner sep=0pt, text width=3cm, minimum height=2cm, font=\bfseries\footnotesize\sffamily},
comment2/.style={rectangle, inner sep= 0pt, text width=3cm, node distance=0.25cm, font=\sffamily}
]
\node[box, fill=red, fill opacity=0.1, color = blue!80] (c2) at (0,0) {};
\node [comment2, left= -3 cm of c2, color = black ] {
$\begin{array}{l l l }
A^2_t = &
\left( 
\begin{array}{l l l }
\sigma^1(t) 	& \ldots 	& \sigma^1(t-60) \\
\ldots  	& \ldots 	& \ldots  \\
\sigma^m(t) 	& \ldots 	& \sigma^m(t-60)
\end{array} \right)
\end{array}
$
};
\end{tikzpicture}
\end{center}

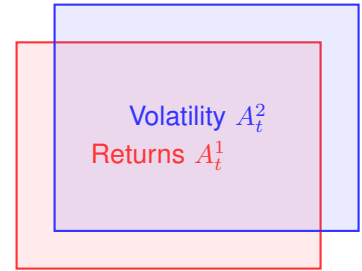
\begin{figure}[h]
\centering
\begin{tikzpicture}
[
comment/.style={rectangle, inner sep= 0pt, text width=3cm, node distance=0.25cm, font=\scriptsize\sffamily},
comment2/.style={rectangle, inner sep= 0pt, text width=3cm, node distance=0.25cm, font=\sffamily}
]

\node[box1, fill=red, fill opacity=0.1, color = red!80] (c2) at (0,0) {};
\node[box1, fill=blue, fill opacity=0.1, color = blue!80] (c1) at (0.5,0.5) { };

\node [comment2, left= -4 cm of c2, color = red!80] {Returns $A^1_t$ };
\node [comment2, left= -4 cm of c1, color =  blue!80] {Volatility $A^2_t$ };
\end{tikzpicture}
\caption{3 dimensional tensor: the asset states}\label{fig:asset_states}
\end{figure}

The asset states are stored in a 3-D tensor as shown in figure \ref{fig:asset_states}. Its similarities with image where pixels are stored in 3 different matrices representing red, green and blue image enable us to use 2 dimensional convolution network for our deep network. The analogy goes even further as it is well known in image recognition that convolutional networks achieves strong performances thanks to their capacity to extract meaningful features and to have very limited parameters hence avoiding over-fitting.

To introduce conceptual based information, the asset manager robot observes also additional important features denoted by $C_t$ that provides insights about the future evolution of the portfolio strategies. Using market knowledge from Homa capital multi assets solutions, we add 3 features (referred to as contextual features) that are correlation between equity and bonds denoted by $c^1_t$, Citigroup global economic surprise index denoted by $c^2_t$, and risk aversion index denoted by $c^3_t$. These features are not taken at random but are well known or at least assumed to have some predictive power for our portfolio strategies as these strategies incorporates a mix of equity and bonds and are highly sensitive to economic surprise and risk aversion level. Again to ensure somehow some Markovianity and to include in the current knowledge of the virtual agent more than the last observation of these features, we introduce a second lag operator $\delta_2$ that operates on the contextual features. To keep things simple in our experience, we take the same vector of lagging periods to represent this second lag operator although the method can be fine-tuned with two different lags for the asset and contextual states. In our setting, $ \delta_2 =  \left[ 0, 1, 2, 3 , 4, 20, 60 \right]$ and the contextual states that is represented by $C_t$ writes as follows:

\begin{center}
\begin{tikzpicture}
[
box/.style={rectangle, draw, fill=red, fill opacity=0.1, color = green!80,
rectangle, minimum height=4cm, minimum width=6cm,
inner sep=0pt, text width=3cm, minimum height=2cm, font=\bfseries\footnotesize\sffamily},
comment2/.style={rectangle, inner sep= 0pt, text width=3cm, node distance=0.25cm, font=\sffamily}
]
\node[box, fill=red, fill opacity=0.1, color = green!80] (c2) at (0,0) {};
\node [comment2, left= -3 cm of c2, color = black ] {
$\begin{array}{l l l }
C_t = &
\left( 
\begin{array}{l l l }
c^1(t) 	& \ldots 	& c^1(t-60) \\
\ldots  	& \ldots 	& \ldots  \\
c^3(t) 	& \ldots 	& c^3(t-60)
\end{array} \right)
\end{array}
$
};
\end{tikzpicture}
\end{center}

In contrast to asset states, contextual states $C_t$ are only represented by a two dimensions tensor or equivalently a matrix. If we want to use convolutional networks, we therefore need to use 1D (for 1 dimensional) and not 2D (2 dimensional) convolutions. In addition, we add in these common contextual features the maximum portfolio strategy's return, the maximum and minimum portfolio strategy's volatilities. The latter two are like for asset states motivated by the stylized fact that standard deviations are useful features to detect crisis.\\

Last but not least we can also introduce that our state $s_t$ incorporates the previous portfolio allocation. Hence our state can take the following three inputs:
\begin{itemize}
\item previous portfolio strategy returns lagged by  $\delta_1$ called the asset states $A_t$;
\item contextual features observed lagged by  $\delta_2$ called the common states $C_t$;
\item the previous weight allocation $w_{t-1}$;
\end{itemize}
or mathematically, $ S_t = \left\{ A_t, C_t, w_{t-1}\right\} $ \\

Our optimal control problem is to find the optimal policy $\pi^*(S_t)$ that maximizes the total reward denoted by $R(T)$ for one episode. Under very strong theoretical assumptions, this optimal policy always exists and is unique. In practice, we are far from the theoretical framework and we may find only locally optimal policies thanks to gradient ascent! The policy is represented by a deep network whose parameters are given by $\theta$ and composed of three sub-networks as illustrated in figure \ref{fig:network} and further described in \ref{sec:system}. Hence the optimal control problem writes as 
$$ \max_{\theta} \mathbf{E}_{\pi_{\theta}(.)} R(T), $$

\noindent where $\mathbf{E}_{\pi_{\theta}}$ represents the expectation under the assumption that our policy $\pi_{\theta}(s_t)$ is precisely represented by our deep networks whose parameters are $\theta$ for a state at time $t$ given by $s_t$.  The total reward $R(T)$ can either be the net performance of the portfolio or some risk return criterion like the Sharpe ratio computed 
as the ratio of the average mean return over its standard deviation. 

\subsection{Network Architecture}\label{sec:system}
\begin{figure}[t]
\centering
\includegraphics[width=\linewidth, height=6.5cm]{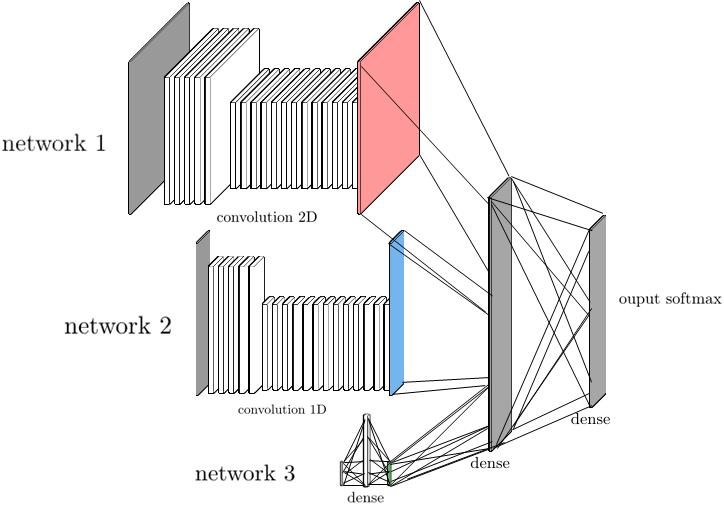}
\caption{Possible DRL network architecture}\label{fig:network}
\end{figure}

Our network (as described in figure \ref{fig:network}) uses three types of inputs:
\begin{itemize}
\item sub-network 1: portfolio returns and standard deviations observed over the lag $\delta_1$ array (the asset states $A_t $); 
\item sub-network 2: contextual information given by the correlation between equities and bonds, the Citigroup economic surprise and the risk aversion indexes observed over the lag $\delta_2$ array and other additional common features like the maximum portfolio strategy's return, the maximum and minimum portfolio strategy's volatilities (the context states $C_t$); 
\item and potentially sub-network 3: the previous portfolio allocation $w_{t-1}$
\end{itemize}

We concatenate these 3 networks into a final one using two dense layers and a final softmax one to infer the portfolio weights. 

Our reward is either the Sharpe ratio or the net value of the final portfolio. In terms of network internal architecture, we can either use convolution layers for sub-network 1 and 2 (convolution 2D and 1D respectively) or LSTM units. We can also do adversarial training by introducing some Gaussian noise in the training to make each iteration slightly different. This helps to have more robust models. 

Concerning the train-validation-test split of our data-set, we use the following split: Train data-set is from 01-Jan-2010 to 31-Dec-2015, validation set from 01-Jan-2016 to 31-Dec-2017, while test data set ranges from 01-Jan-2018 to 31-Mar-2020. Hyper-parameters are tested on the validation sets. We provide the hyper-parameters used in the final run in table  \ref{tab:hyperparam}. Results are quite sensitive to the Adam learning rate and the lag 1 and 2 arrays. We tried various solutions and found that taking the last week of observation, the last month and the last quarter was working well and quite intuitive for the the lag 1 and 2 arrays. Results of the various trained networks and performance over iterations can be visualized in  \href{http://www.aisquareconnect.com/deeprl/ICPRSummary.html}{http://www.aisquareconnect.com/deeprl/ICPRSummary.html} and are also given as supplementary materials of this paper. \\

\begin{figure}[t]
\centering
\includegraphics[width=\linewidth]{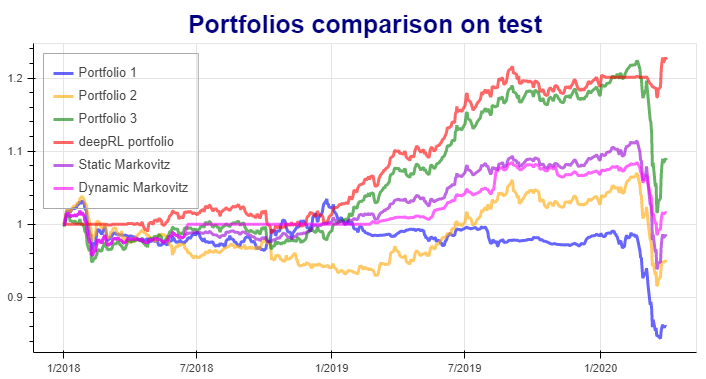}
\caption{Deep RL Portfolio Optimisation Result}\label{fig:backtest_best}
\end{figure}

All in all, the different possible network configurations and architectures represent 32 models whose results are given in table  \ref{tab:results}. In our experiment, $m=4$ with the first three assets representing real strategies, while the fourth one being just cash whose value do not change over time. To represent the performance of each of the 3 strategies, we plot portfolio 1 which consists in taking only strategy 1 (in blue in figure \ref{fig:backtest_best}), respectively portfolio 2 and 3 taking only strategy 2 and 3 (in orange and green). 

\begin{figure}[t]
\centering
\includegraphics[width=\linewidth]{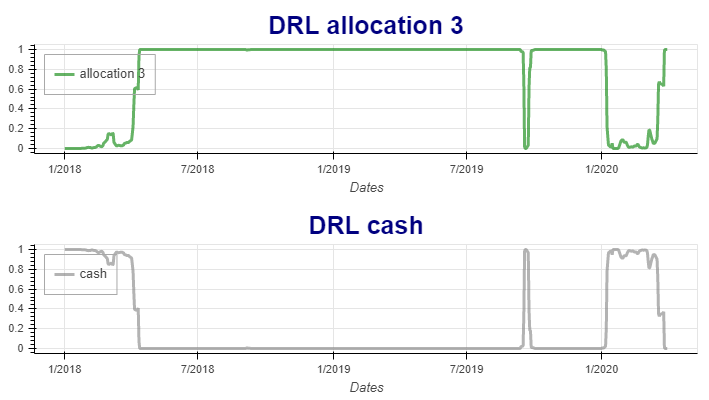}
\caption{Corresponding optimal allocation}\label{fig:backtest_best_weights}
\end{figure}

It is worth noticing that the portfolio 3 consisting of 100 \% in strategy 3 has a strong tendency to over-perform the other two strategies (portfolio 1 and 2). Hence we expect the deep RL agent to allocate mostly in strategy 3 and when anticipating a crisis, to allocate in cash. This is exactly what it does as illustrated in figure \ref{fig:backtest_best_weights}. It is also interesting to notice that the trained deep RL agent is mostly invested in strategy 3 and from time to time swap this allocation to a pure cash allocation. The anticipated crisis in 2018 enables the agent to slightly over-perform portfolio 3 from 2018 to the end of 2019. The agent however is not all mighty and makes mistake as illustrated by the wrong peaked cash allocation in end of 2019. It is able to adapt to the Covid crisis and to brutally swap allocation from strategy 3 to cash and back as markets bounced back at the end of March.

\subsection{DRL algorithm}
To find the optimal action $\pi^*(S_t)$ (in terms of portfolio allocation), we use deep policy gradient method with non linear activation (Relu). We use buffer replay to memorize all marginal rewards, so that we can start batch gradient descent once we reached the final time step. We use the traditional Adam optimization so that we have the benefit of adaptive gradient descent with root mean square propagation \cite{kingma2014method}. 

\subsection{Results}
Performance results are given below in table \ref{tab:results}. Best performing models are highlighted in yellow. Returns are computed annually. Hence for a total performance of 21 \% (as shown in figure \ref{fig:backtest_best}) over the period of January 1st 2018 to March 31st 2020, the corresponding annual return is 8.8 \%. Overall, out of the 32 models available, there are many DRL models that are able to over-perform not only traditional methods like static Markovitz but also the best portfolio (sometimes referred as the naive winner strategy) in terms of net performance and Sharpe ratio, with a final annual net return of 8.8\% when using the best net profit reward model or 8.6\% when using the best Sharpe ratio reward model compared to 3.9 \% for the naive winner method. Dynamic Markovitz method consists in computing the Markovitz optimal allocation every 3 months. The Naive winner method consists in just selecting the best strategy over the train data set, which is strategy 3.

{\small
\begin{table}[htbp]
  \centering
  \caption{Performance results} \label{tab:results0}
  	\resizebox{0.5 \textwidth} {!} {
    \begin{tabular}{|c|c|c|c|c|c|c|c|}
	\hline
	\rule[-0.3cm]{0cm}{0.8cm}
          & Portfolio &  Portfolio & Portfolio & Dynamic		& Deep RL 		& Deep RL 	& Naive \\
          &  1 	 &  2		   & 3 		& Markovitz 	& Net\_profit 	& Sharpe   & winner \\
	\hline
	\rule[-0.3cm]{0cm}{0.8cm}
        Net Performance & -6.3\% & -2.1\% & 3.9\%  & 0.7\%  & \textbf{8.8\%} & 8.6\%  & 3.9\% \\
	\hline
	\rule[-0.3cm]{0cm}{0.8cm}
	Std dev & 6.1\%  & 6.5\%  & 7.3\%  & 4.3\%  & 4.5\%  & 4.2\%  & 7.3\% \\
	\hline
	\rule[-0.3cm]{0cm}{0.8cm}
	Sharpe ratio & na     & na     &                    0.53  &                         0.17  &                                1.95  & \textbf{2.08} &                    0.53  \\
	\hline
    \end{tabular}
}
\end{table}
}

\section{Learning of the network parameters}
The agent's objective is to maximize its total reward $R$ given at episode end. This reward can be net portfolio performance or Sharpe ratio computed as portfolio mean return over its standard deviation. Because we somehow play and play again the same scenario with the same reward function, the current framework has two important distinctions from many other RL problems. One is that the domain knowledge of the environment is well-mastered, and can be fully exploited by the agent. 
This exact expressiveness is a direct consequence that the agent's action has no influence on future price, which is clearly the case for small transactions or liquid assets. This isolation 
of action and external environment also allows one to use the same segment of market history to evaluate difference sequences of actions. 

The second distinction is that the final reward depend on all episodic actions. In other words all episodic actions are important, justifiying the full-exploitation approach.

\subsection{Deterministic Policy Gradient}
A policy is a mapping from the state space to the action space, $\pi:\mathcal{S}\rightarrow\mathcal{A}$. With full exploitation in the current framework, an action is deterministically produced by the policy from a state. The optimal policy is obtained using a gradient ascent algorithm. To achieve this, a policy is specified by a set of parameter $\vec \theta$, and $\vec a_t = \pi_{\vec \theta}(\bm s_t)$. The performance
metric of $\pi_{\vec \theta}$ for time interval $[0,t]$ is defined as the corresponding reward function of the interval,
\begin{equation}
	J_{[0,t]}(\pi_{\vec \theta}) = R\left( \vec s_1,\pi_{\vec \theta}(s_1),\cdots,
		\vec s_{t},\pi_{\vec \theta}(s_{t}),\vec s_{t+1} \right).
	\label{eq:policy_value}
\end{equation}
After random initialization, the parameters are continuously updated along the gradient direction with a learning rate $\lambda$,
\begin{equation}
	\vec\theta \longrightarrow \vec\theta + \lambda\nabla_{\vec\theta}J_{[0,t]}(\pi_{\vec \theta}).
	\label{eq:gradient_ascent}
\end{equation}
To make the gradient ascent optimization, we use the standard Adam (short for Adaptive Moment Estimation) optimizer to have the benefit of adaptive gradient descent with root mean square propagation \cite{kingma2014method}.

\subsection{Crisis adaptation}
\begin{figure}[t]
\centering
\includegraphics[width=\linewidth]{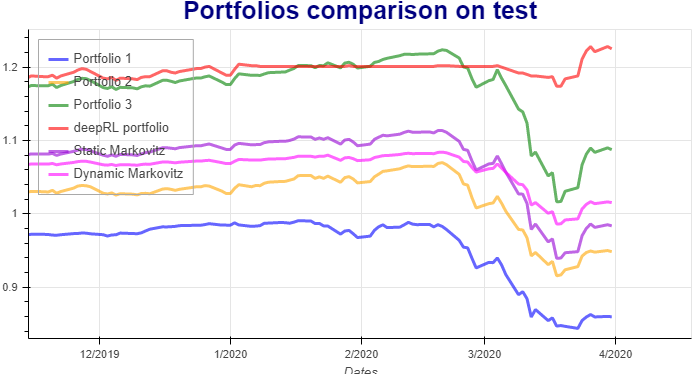}
\caption{Crisis portfolio reaction. It is worth noticing that the best DRL agent has a stable performance as it rapidly dis-invests and put asset in cash during the Covid crisis}\label{fig:crisis_1}
\end{figure}

\begin{figure}[t]
\centering
\includegraphics[width=\linewidth]{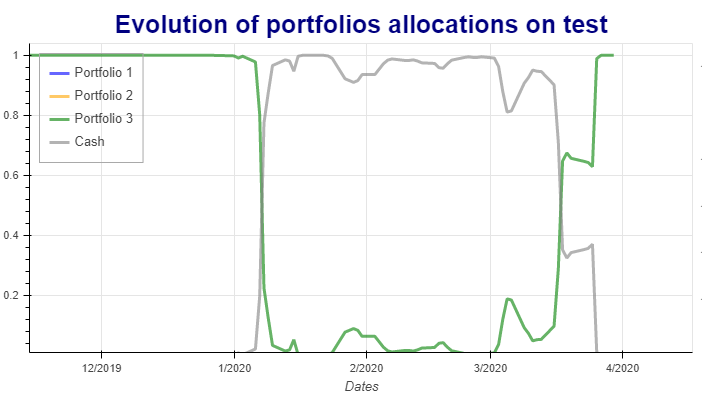}
\caption{Allocation during Covid crisis. It is worth noticing that the best DRL agent  allocates almost all asset either in cash or strategy 3 and is rarely mixing the two strategies. It does not include at all strategies 1 or 2, indicating that the optimal choice is to saturate the allocation constraints as we only permit the dynamic agent to allocate between 0 and 100 \%}\label{fig:crisis_2}
\end{figure}

It is remarkable that the DRL approach is able to handle the Covid crisis softly as displayed by figure \ref{fig:crisis_1}. If we zoom over the period out of sample from December 2019 to March 2020, we can see that the DRL agent is able to rapidly reduce exposure to strategy 3 and allocate in cash, detecting thanks to contextual information that a crisis is imminent as show in figure \ref{fig:crisis_2}.  Interestingly, the DRL agent reallocates to portfolio 3  in March picking the market rebound.

\begin{figure}[t]
\centering
\includegraphics[width=0.9 \linewidth]{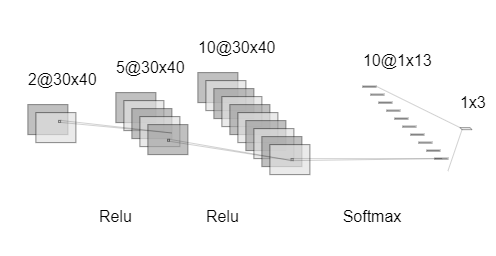}
\caption{convolutional network 1}\label{fig:best_network}
\end{figure}

Best networks as illustrated in table \ref{tab:results} are mostly convolutional networks. We found that for the sub network 1, it is optimal to have 2 convolutional layers but only 1 convolutional layer for the sub network 2. We illustrate the sub-network 1 in figure \ref{fig:best_network}. 

\subsection{Impact of contextual information}
Logically, networks with contextual information performs better as they have more information. For each network configuration, we compute the difference between the version with and without contextual information. We summarized these results in table \ref{tab:results2} with best results highlighted in yellow. For all configurations, the version with contextual information achieves higher annual returns. This is almost the case also on Sharpe ratio, but there are exceptions. If we remove for each criterion the two largest difference classified as outlyers, we found that contextual based models increase on average annual returns by 2.45 \% and Sharpe ratio by 0.29.

\section{Further work}
On experiment, we see that the contextual based approach over-performs baseline methods like Markovitz. 
We also experienced that CNN architecture performs much better than LTSM units as they reduce the number of parameters to train and share parameters across portfolio strategies.
Adversarial training makes also the training more robust by providing a more challenging environment. Last but not least, it is quite important to fine tune the numerous hyper-parameters of the contextual based DRL model, namely the various lags (lags period for the sub network fed by portfolio strategies past returns, lags period for common contextual features referred to as the common features in the paper), standard deviation period, learning rate, etc... 
It is compelling that the suggested framework is linearly scalable with the portfolio size and can accommodate contextual information.
Our findings suggest that modeling the state with previous weight allocation deteriorates training and does not help suggesting that the artifact of introducing previous weight to have a direct impact on state when performing an action is artificial and that in reality under the assumption of small market impact, it is more efficient to assume that portfolio allocation does not influence future state. 
The memory mechanism is quite beneficial as it allows to compute the final reward on each episode and hence allows avoiding gradient vanishing problem faced by many deep networks. 
Moreover, thanks to this memory mechanism, it is not challenging to create an online learning mechanism that can continuously digest incoming market information to improve the dynamic agent.
The profitability of this framework surpasses traditional portfolio-selection methods, as demonstrated in the paper by a non negligible factor as it outperforms dynamic Markovitz by than 8 \% and the best strategy by 4 \%. 

This better performance should be mitigated by the fact that the dynamic DRL agent is able to adapt well to the Covid crisis. Hence it benefits from an exceptional and almost unique condition in the financial history. Consequently, these numbers should not be taken literally but rather as a sign of the capacity of deep RL method to achieve human performance in portfolio allocation and to be able to detect and adapt to crisis patterns. Despite the efficiency of contextual based DRL models in experiments, these models can be improved in future works.  Their main weakness is the number of hyper parameters that needs to be estimated on the validation set. Their second major weakness relies on the fact that in finance, each experience is somehow unique and one may not be able to draw conclusion on a single test set. Drawing a general conclusion is premature and beyond reason at this stage. It should be tested on more financial markets and on more outcomes. It may also be tested in terms of stability and capacity to adapt to further crisis patterns. 

\section{Conclusion}\label{sec:conclusion}
In this paper, we address the challenging task of detecting and adapting portfolio allocation to crisis environment.
Our approach is based on deep reinforcement learning using contextual information thanks to a second sub-network.
The model takes not only past performances of portfolio strategies over different rolling period, but also portfolio strategies standard deviation as well as contextual information like risk aversion, Citigroup economic surprise index, correlation between equity and bonds over a rolling period to make best allocation decision.
The additional contextual information makes the learning of the dynamic asset manager agent more robust to crisis environment as the agent reacts more rapidly to changing environments. 
In addition, the usage of standard deviation of portfolio strategies provides a good hint for future crisis. The model achieves better performance than standard financial models. There are room for further improvement as this model constitutes only a first attempt to find a reasonable DRL solution to adapt to crisis situation and to answer positively if DRL can reach human level in applications to financial problems and in particular in detecting pattern crisis.

\begin{table}[htbp]
  \centering
  \caption{Results of the various models}
\resizebox{0.50 \textwidth} {!} {
    \begin{tabular}{|l|c|l|c|c|r|r|}
    \toprule
    Reward & Adversarial 	& Network & Previous & Context?	&  Annual & Sharpe \\
		& training?   	&		& weight?	&		& return 		& \\
    \midrule
    NetProfit & No  & Conv2D & No  & Yes   & \colorbox{yellow}{\textbf{8.8\%}} &        1.95  \\
    \midrule
    Sharpe & Yes & Conv2D & No  & Yes & 8.6\% & \colorbox{yellow}{\textbf{ 2.08 }} \\
    \midrule
    NetProfit & No  & Conv2D & Yes   & Yes  & 8.5\%  &         2.03  \\
    \midrule
    Sharpe & No  & Conv2D & No  & Yes   & 8.4\%  &         2.01  \\
    \midrule
    NetProfit & Yes   & Conv2D & No  & Yes   & 8.0\%  &         1.35  \\
    \midrule
    NetProfit & Yes   & Conv2D & No  & No  & 7.7\%  &         1.94  \\
    \midrule
    Sharpe & No  & Conv2D & No  & No  & 6.4\%  &         1.31  \\
    \midrule
    NetProfit & No  & LSTM   & No  & Yes   & 6.2\%  &         1.49  \\
    \midrule
    NetProfit & No  & Conv2D & No  & No  & 5.4\%  &         0.97  \\
    \midrule
    Sharpe & Yes   & LSTM   & No  & Yes   & 5.4\%  &         1.23  \\
    \midrule
    NetProfit & Yes   & LSTM   & No  & Yes   & 5.1\%  &         0.93  \\
    \midrule
    NetProfit & Yes   & Conv2D & Yes   & Yes   & 4.3\%  &         0.63  \\
    \midrule
    Sharpe & Yes   & Conv2D & No  & No  & 4.2\%  &         0.69  \\
    \midrule
    NetProfit & No  & LSTM   & Yes   & Yes   & 3.8\%  &         0.52  \\
    \midrule
    Sharpe & No  & Conv2D & Yes   & No  & 3.8\%  &         0.52  \\
    \midrule
    NetProfit & No  & Conv2D & Yes   & Yes   & 3.8\%  &         0.52  \\
    \midrule
    Sharpe & Yes   & LSTM   & Yes   & Yes   & 3.8\%  &         0.52  \\
    \midrule
    Sharpe & Yes   & Conv2D & Yes   & Yes   & 3.7\%  &         0.51  \\
    \midrule
    NetProfit & Yes   & Conv2D & Yes   & No  & 3.7\%  &         0.51  \\
    \midrule
    NetProfit & No  & LSTM   & Yes   & No  & 3.6\%  &         0.49  \\
    \midrule
    NetProfit & Yes   & LSTM   & Yes   & Yes   & 3.5\%  &         0.48  \\
    \midrule
    NetProfit & Yes   & LSTM   & No  & No  & 3.4\%  &         1.24  \\
    \midrule
    Sharpe & No  & LSTM   & Yes   & Yes   & 3.4\%  &         0.48  \\
    \midrule
    NetProfit & No  & LSTM   & No  & No  & 3.4\%  &         0.47  \\
    \midrule
    Sharpe & Yes   & Conv2D & Yes   & No  & 3.4\%  &         0.51  \\
    \midrule
    NetProfit & Yes   & LSTM   & Yes   & No  & 2.3\%  &         0.97  \\
    \midrule
    Sharpe & Yes   & LSTM   & No  & No  & 2.3\%  &         0.32  \\
    \midrule
    Sharpe & No  & LSTM   & No  & Yes   & 1.5\%  &         0.22  \\
    \midrule
    Sharpe & No  & Conv2D & Yes   & No  & 0.9\%  &         0.13  \\
    \midrule
    Sharpe & Yes   & LSTM   & Yes   & No  & -5.1\% & na \\
    \midrule
    Sharpe & No  & LSTM   & No  & No  & -5.1\% & na \\
    \midrule
    Sharpe & No  & LSTM   & Yes   & No  & -5.1\% & na \\
    \bottomrule
    \end{tabular}%
}
  \label{tab:results}%
\end{table}%

\begin{table}[htbp]
  \centering
  \caption{Difference in returns and sharpe between model with and without contextual information}
\resizebox{0.50 \textwidth} {!} {
    \begin{tabular}{|l|c|l|c|r|r|}
    \toprule
    Reward & Adversarial 	& Network & Previous &Annual return   & Sharpe \\
		& training?   	&		& weight?			& 	difference	& difference \\
    \midrule
    NetProfit & No  & Conv2D & No  & 3.43\% &           0.99  \\
    \midrule
    NetProfit & No  & Conv2D & Yes   & 4.68\% &           \colorbox{yellow}{\textbf{ 1.51}}  \\
    \midrule
    NetProfit & No  & LSTM   & No  & 2.77\% &           1.03  \\
    \midrule
    NetProfit & No  & LSTM   & Yes   & 0.27\% &           0.03  \\
    \midrule
    NetProfit & Yes   & Conv2D & No  & 0.32\% & -        0.59  \\
    \midrule
    NetProfit & Yes   & Conv2D & Yes   & 0.58\% &           0.12  \\
    \midrule
    NetProfit & Yes   & LSTM   & No  & 1.70\% & -        0.32  \\
    \midrule
    NetProfit & Yes   & LSTM   & Yes   & 1.16\% & -        0.48  \\
    \midrule
    Sharpe & No  & Conv2D & No  & 2.02\% &           0.70  \\
    \midrule
    Sharpe & No  & Conv2D & Yes   & 2.94\% &           0.40  \\
    \midrule
    Sharpe & No  & LSTM   & No  & 6.56\% &           0.22  \\
    \midrule
    Sharpe & No  & LSTM   & Yes   & 8.52\% &           0.48  \\
    \midrule
    Sharpe & Yes   & Conv2D & No  & 4.39\% &           1.39  \\
    \midrule
    Sharpe & Yes   & Conv2D & Yes   & 0.36\% &           0.01  \\
    \midrule
    Sharpe & Yes   & LSTM   & No  & 3.05\% &           0.91  \\
    \midrule
    Sharpe & Yes   & LSTM   & Yes   &  \colorbox{yellow}{\textbf{8.87\%}} &           0.52  \\
    \bottomrule
    \end{tabular}%
}
  \label{tab:results2}%
\end{table}%

\begin{table}[htbp]
  \centering
  \caption{Hyper parameters used}
\resizebox{0.50 \textwidth} {!} {
    \begin{tabular}{|p{6em}|p{3.2em}|p{17em}|}
    \toprule
    hyper-parameters & value  & description \\
    \midrule
    batch size & \multicolumn{1}{r|}{50} & Size of mini-batch during training \\
    \midrule
    regularization coefficient & \multicolumn{1}{r|}{1e-8}   & $L_2$ regularization coefficient applied to network training \\
    \midrule
    learning rate & \multicolumn{1}{r|}{0.01} & Step size parameter in Adam \\
    \midrule
    standard deviation period & \multicolumn{1}{r|}{20 days} & period for standard deviation in asset states \\
    \midrule
    commission & \multicolumn{1}{r|}{10 bps} & commission rate  \\
    \midrule
    stride & \multicolumn{1}{r|}{2,1} & stride used in convolution networks \\
    \midrule
    conv  number 1& \multicolumn{1}{r|}{5,10 } & number of convolutions in sub-network 1\\
    \midrule
    conv  number 2& \multicolumn{1}{r|}{2} & number of convolutions in sub-network 2\\
    \midrule
    lag period 1 & \multicolumn{1}{r|}{$\left[ 60, 20, 4,3,2,1,0 \right]$ } & lag period for asset states\\
    \midrule
    lag period 2 & \multicolumn{1}{r|}{$\left[ 60, 20, 4,3,2,1,0 \right] $} & lag period for contextual states\\
    \midrule
     noise & \multicolumn{1}{r|}{0.002} & adversarial Gaussian standard deviation \\
    \bottomrule
    \end{tabular}%
}
  \label{tab:hyperparam}%
\end{table}%

\bibliographystyle{IEEEtran}
\bibliography{main}
\end{document}